\documentclass[aps,prl,twocolumn,secnumarabic,amssymb,nobibnotes,superscriptaddress]{revtex4-1}
\usepackage{amsmath,amssymb,graphicx,color,bm,soul}
\usepackage{ulem}
\begin{document}

\title{Creation and manipulation of stable dark solitons and \\ vortices in microcavity polariton condensates}% Force line breaks with \\

%\author{Xuekai Ma$^{1,*}$, Oleg A. Egorov$^2$, and Stefan Schumacher$^1$}
%\affiliation{$^1$Department of Physics and Center for Optoelectronics and Photonics Paderborn (CeOPP), Universit\"{a}t Paderborn, Warburger Strasse 100, 33098 Paderborn, Germany}
%\affiliation{$^2$Institute of Condensed Matter Theory and Solid State Optics, Abbe Center of Photonics, Friedrich-Schiller-Universit\"{a}t Jena, Max-Wien-Platz 1, 07743 Jena, Germany}

\author{Xuekai Ma}
\email{Corresponding author: xuekai.ma@gmail.com}
\affiliation{Department of Physics and Center for Optoelectronics and Photonics Paderborn (CeOPP), Universit\"{a}t Paderborn, Warburger Strasse 100, 33098 Paderborn, Germany}

\author{Oleg A. Egorov}
\affiliation{Institute of Condensed Matter Theory and Solid State Optics, Abbe Center of Photonics, Friedrich-Schiller-Universit\"{a}t Jena, Max-Wien-Platz 1, 07743 Jena, Germany}

\author{Stefan Schumacher}
\affiliation{Department of Physics and Center for Optoelectronics and Photonics Paderborn (CeOPP), Universit\"{a}t Paderborn, Warburger Strasse 100, 33098 Paderborn, Germany}

\begin{abstract}
Solitons and vortices obtain widespread attention in different physical systems as they offer potential use in information storage, processing, and communication. In exciton-polariton condensates in semiconductor microcavities, solitons and vortices can be created optically. However, dark solitons are unstable and vortices cannot be spatially controlled. In the present work we demonstrate the existence of stable dark solitons and vortices under non-resonant incoherent excitation of a polariton condensate with a simple spatially periodic pump. In one dimension, we show that an additional coherent light pulse can be used to create or destroy a dark soliton in a controlled manner. In two dimensions we demonstrate that a coherent light beam can be used to move a vortex to a specific position on the lattice or be set into motion by simply switching the periodic pump structure from two-dimensional (lattice) to one-dimensional (stripes). Our theoretical results open up exciting possibilities for optical on-demand generation and control of dark solitons and vortices in polariton condensates.
\end{abstract}

%\pacs{71.36.+c, 03.75.Kk, 42.65.Sf, 71.35.Gg }

\maketitle

Over the past decade, exciton-polaritons in quantum-well semiconductor microcavities have attracted a lot of attention due to their exceptional properties. Exciton-polaritons are quasi particles formed from a cavity light field and a quantum well (QW) exciton \cite{Weisbuch-prl-1992}. These quasi particles possess both photonic and excitonic nature. They have a small effective mass ($\sim10^{-4}m_e$, $m_e$ is the free electron mass) and short lifetime on a picosecond time scale both inherited from the photonic part. Through their excitonic part they interact with each other and as composite Bosons they can undergo a non-equilibrium phase transition with similarities to Bose-Einstein condensation (BEC) \cite{Deng-science-2002,Kasprzak-nature-2006,Deng-RevModPhys-2010}, potentially even up to room temperature \cite{Christopoulos-2007-prl,Christmann-2008-apl,Baumberg-2008-prl}. The interaction between polaritons leads to an effective optical nonlinearity that has led to the observation of a whole wealth of nonlinear phenomena some of which may find applications in functional polariton devices, including bistability \cite{Tredicucci-pra-1996,Baas-pra-2004,Bajoni-prl-2008,Goblot-prl-2016}, pattern formation \cite{Borgh-prb-2010,Luk-prb-2013,Werner-prb-2014,Liew-prb-2015}, vortices \cite{Lagoudakis-NatPhys-2008,Lagoudakis-Science-2009,Roumpos-NatPhys-2011}, and solitons \cite{Egorov-2009-prl,Sich-2012-nph,Yulin-2008-pra,Grosso-2012-prb,Cilibrizzi-2014-prl}.

Solitons and vortices are spatially localized stationary solutions in a nonlinear system. They are investigated in a large variety of physical systems, such as in nonlinear optics \cite{Kivshar-2003-book}, atomic condensates \cite{Dum-1998-prl}, liquid helium \cite{Williams-1999-prl}, superconductors \cite{Harada-1996-Science,Roditchev-2015-nphy}, and magnetic systems \cite{Chen-1993-prl,Shinjo-2000-Science}. Generally speaking, only a focusing nonlinearity (corresponding to attractive interaction in a particle system) can support bright solitons, while dark solitons, which are low-density defects in the homogeneous phase, can only exist in a defocusing (corresponding to repulsive interaction) nonlinear system. In semiconductor microcavities, under optically resonant (coherent) excitation and using the non-parabolicity of the lower polariton branch, for polaritons in the same polarization state such that the nonlinearity is defocussing, the existence of both bright \cite{Egorov-2009-prl,Sich-2012-nph} and dark \cite{Yulin-2008-pra,Grosso-2012-prb,Cilibrizzi-2014-prl} solitons propagating with a finite momentum was reported. Under non-resonant (incoherent) excitation, 2D phase defects (vortices) of polariton condensates are found to be stable \cite{Liew-prb-2015}. However, during the condensation process these entities are randomly created and they cannot be controlled. For the spinor system in 1D, the existence of stable dark soliton trains has been reported \cite{Pinsker-2014-prl}. In the scalar system, dark solitons are unstable and can only persist for some time but then disappear as the system transitions into the stable homogeneous phase in 1D \cite{Xue-2014-prl}. In 2D unstable dark solitons split up into several vortex-antivortex pairs \cite{Smirnov-2014-prb}.

In the present work, we study polariton condensates under non-resonant excitation with a periodic pump profile. We demonstrate the existence, controlled creation and annihilation, and manipulation of stable dark solitons. Besides dark solitons, fully periodic solutions can also be generated. In these dark solitons can then be created and annihilated at desired spatial positions by short coherent light pulses. Such flexible on demand creation of dark solitons has not yet been achieved in other physical systems. We also find stable phase defects for periodic excitation in 2D, so-called vortices. For vortices, different approaches to control their topological charge have been discussed and implemented in the past in spatially structured setups \cite{Miyahara-1985-apl,Cowburn-2007-nmat,Sigurdsson-prb-2014,Ma-prb-2016}. The vortices in the present work are pinned to their initial position on the optically imprinted 2D lattice. However, we demonstrate that a coherent light beam at $k=0$ can help a vortex escape its potential trap such that it is free to move to an adjacent cell, following the motion of the coherent beam on the 2D lattice. Additionally, the vortices can also be controlled switching the non-resonant periodic pump from a 2D lattice to stripes, setting already existing vortices into motion along the stripes.

\begin{figure} %[htbp]
\includegraphics[width=0.95\columnwidth]{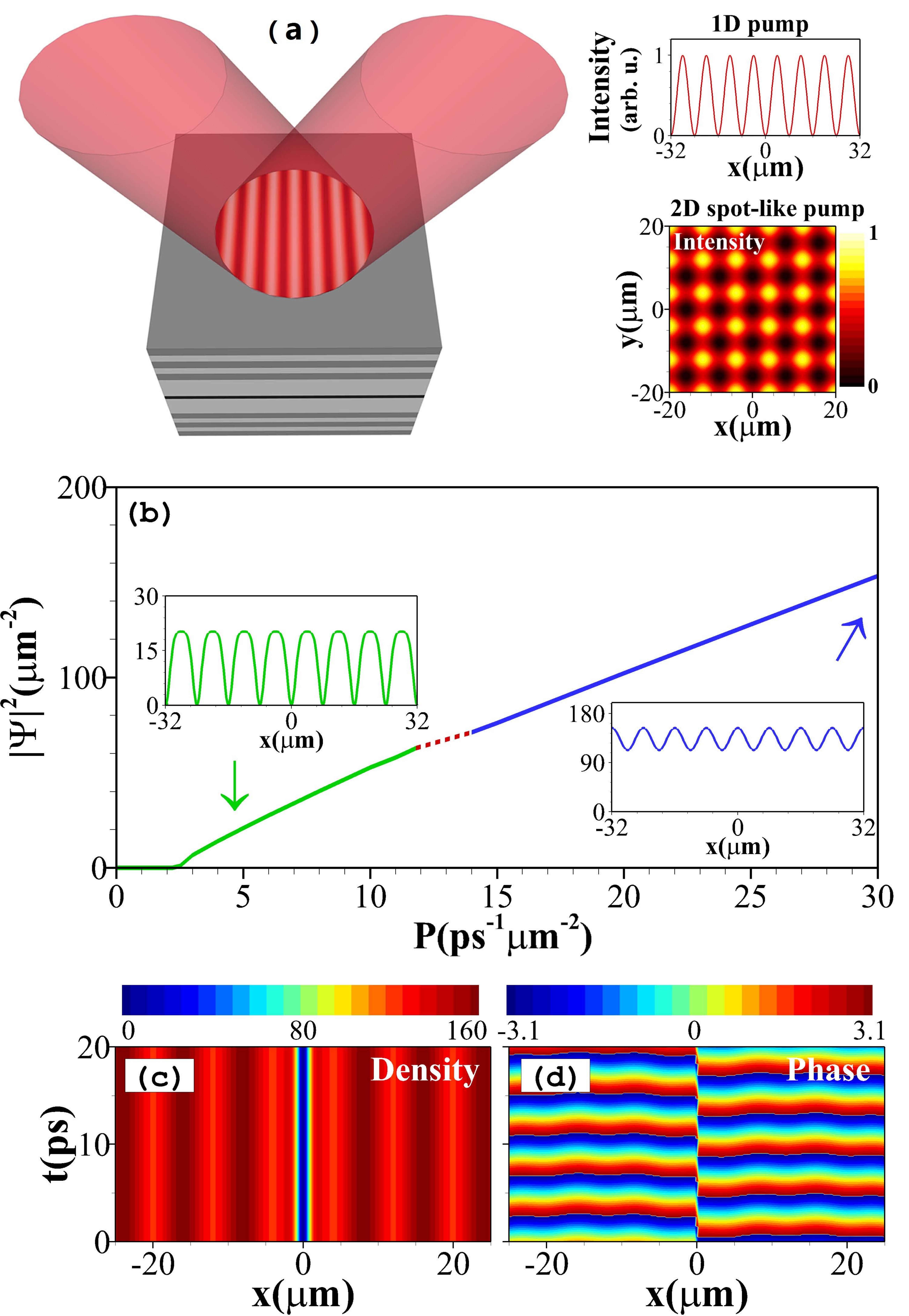}
\caption{(Color online) (a) Sketch of a semiconductor microcavity. Two coherent homogenous optical beams create a periodic reservoir. The two panels on the right illustrate the periodic pump profiles in 1D and 2D. The 2D periodic structure is generated by four coherent homogeneous beams. (b) Density distribution of steady state periodic solutions created by periodic pumps in 1D. Solid lines represent stable solutions, while the dashed line represents unstable solutions. The insets show the solutions at $P=5$ and $P=30$, respectively. Time evolution of (c) the density and (d) the phase of a stable dark soliton in 1D at $P=30$.}\label{f1}
\end{figure}

\textit{Model} -- To study the dynamics of a polariton condensate, a driven-dissipative Gross-Pitaevskii (GP) model can be used to describe the condensate dynamics coupled to an exciton reservoir \cite{Wouters-prl-2007}:
%\begin{widetext}
\begin{equation}\label{e1}
\begin{aligned}
i\hbar\frac{\partial\Psi(\mathbf{r},t)}{\partial t}&=\left[-\frac{\hbar^2}{2m}\nabla_\bot^2-i\hbar\frac{\gamma_c}{2}+g_c|\Psi(\mathbf{r},t)|^2 \right.\\
&+\left.\left(g_r+i\hbar\frac{R}{2}\right)n(\mathbf{r},t)\right]\Psi(\mathbf{r},t)+P_{c}(\mathbf{r},t)\,,
\end{aligned}
\end{equation}
\begin{equation}\label{e2}
\frac{\partial n(\mathbf{r},t)}{\partial t}=\left[-\gamma_r-R|\Psi(\mathbf{r},t)|^2\right]n(\mathbf{r},t)+P_{i}(\mathbf{r},t)\,.
\end{equation}
%\end{widetext}
Here $\Psi(\mathbf{r},t)$ is the coherent polariton field and $n(\mathbf{r},t)$ is the exciton reservoir density. $m=10^{-4}m_e/a$ is the effective mass of polaritons on the lower branch, with a parameter $a$ which can be adjusted to model different detunings of the cavity mode from the quantum well exciton resonance. $\gamma_c$ and $\gamma_r=1.5\gamma_c$ are the loss rates of polaritons and reservoir, respectively, $R=0.01\,\mathrm{ps^{-1}\mu m^2}$ is the condensation rate, $g_c=6\times10^{-3}\,\mathrm{meV\mu m^2}$ represents the nonlinear interaction between polaritons, and $g_r=2g_c$ is the interaction between polaritons and reservoir excitons \cite{Roumpos-NatPhys-2011}. $P_c(\mathbf{r},t)$ is a coherent pulse. The periodical incoherent pump, $P_i(\mathbf{r},t)$, is shown in Fig.~\ref{f1}(a). The one- and two-dimensional periodic pump structures are generated by interfering two (or four, respectively) coherent plane waves as illustrated.

\textit{1D excitation} -- First we study a one-dimensional system where polaritons are confined in a wire in the cavity plane. Such a structure can be realized using a number of different techniques \cite{Wertz-nph-2010,Kaitouni-prb-2006,Cerda-prl-2010,Lai-nature-2007,Kim-np-2011,Tanese-nc-2013,Baboux-prl-2016,Winkler-njp-2015,Winkler-prb-2016}. The periodic pump source is given by $P_i(x)=P\sin^2(\pi x/d)$ with the period $d=8\,\mathrm{\mu m}$ as illustrated in Fig.~\ref{f1}(a). For a continuous wave source, Fig.~\ref{f1}(b) shows the region of existence of periodic stationary solutions depending on the pump intensity $P$. When the pump intensity is above condensation threshold and not too high with $2\lesssim P\lesssim12$, the density distribution of the periodic solution is similar to the periodic reservoir density $n(x)$ and pump distribution. That is, the minimum density of the condensate solution resides in the valleys of the pump density at $x=d\cdot N$, $N=0,\pm1,\pm2,\ldots$ as shown in the inset of Fig.~\ref{f1}(b). However, when $P>14$, the periodic reservoir density acts as a periodic potential trapping condensate density inside the valleys such that the maximum condensate density is found in the valleys of the pump at $x=d\cdot N+d/2$. In the small region of $12\lesssim P\lesssim14$ no stable stationary solutions are found. Interestingly, we find that in the region $P>14$, besides the periodic solutions, also stable dark solitons form in one or multiple of the pump valleys. Figure~\ref{f1}(c) shows an example with a dark soliton at $x=0$. The drop in density is accompanied by a $\pi$ shift in phase.

\begin{figure} %[htbp]
\includegraphics[width=1.0\columnwidth]{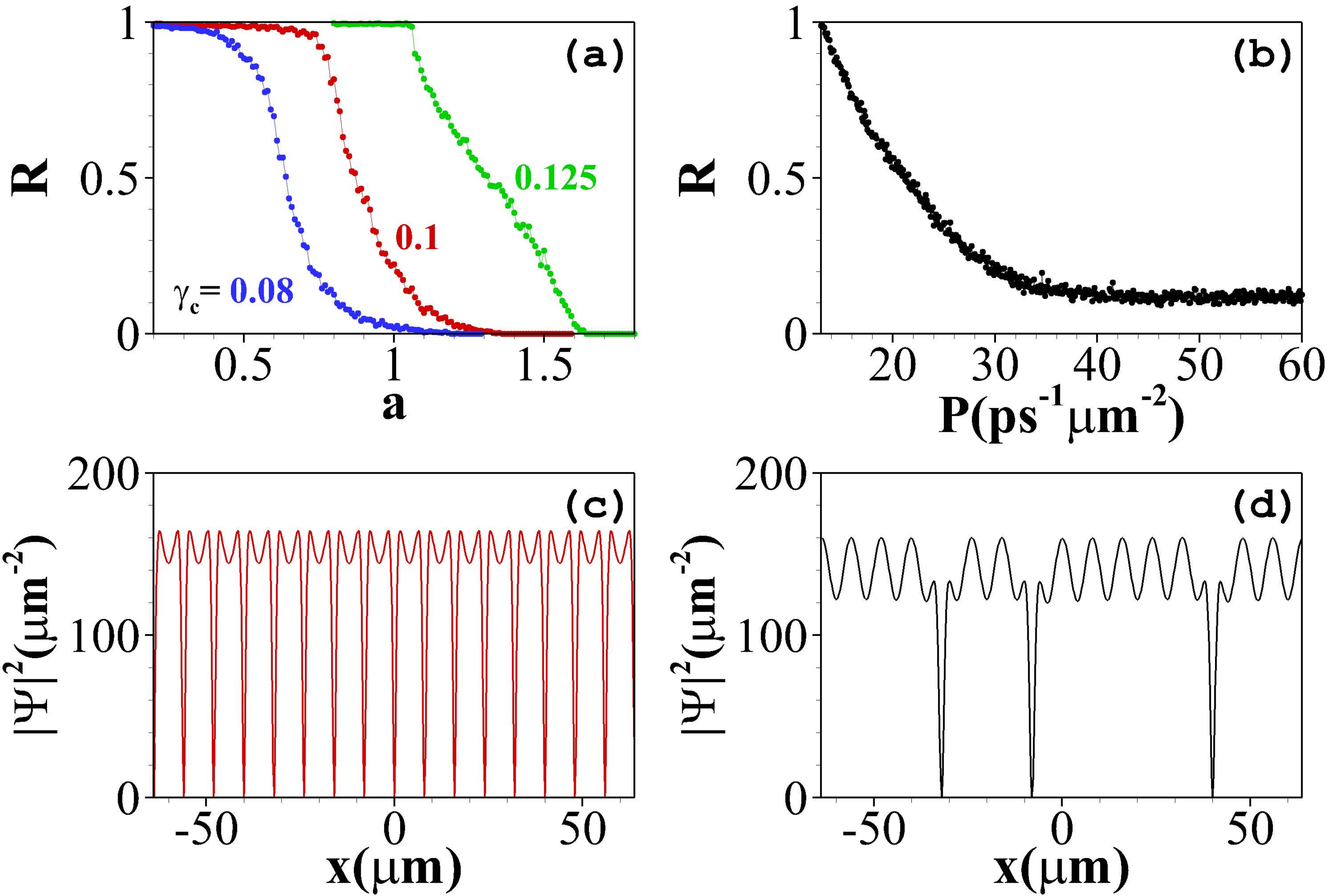}
\caption{(Color online) Density of dark solitons for different system parameters. (a) Dependence of the number of dark solitons per number of pump valleys, $R$, on the effective polariton mass (scaled with the factor $a$) at $P=30$ for $\gamma_c=0.08\,\mathrm{ps^{-1}}$, $\gamma_c=0.1\,\mathrm{ps^{-1}}$, and $\gamma_c=0.125\,\mathrm{ps^{-1}}$. (b) Dependence of $R$ on the pump intensity with $a=1$ and $\gamma_c=0.1\,\mathrm{ps^{-1}}$. Profiles of dark solitons with (c) $\gamma_c=0.1\,\mathrm{ps^{-1}}$, $a=0.5$, and $P=30$, and with (d) $\gamma_c=0.1\,\mathrm{ps^{-1}}$, $a=1$, and $P=30$.}\label{f2}
\end{figure}

In the real system, the onset of condensate formation occurs spontaneously from noise resulting in a certain number of dark solitons. The system parameters influence the robustness of initial phase defects. To quantify the number of dark solitons, we define the ratio $R=\frac{\text{number of dark solitons}}{\text{number of pump valleys}}$, with $0\leq R\leq1$. Figure~\ref{f2}(a) illustrates how the number of dark solitons depends on polariton loss $\gamma_c$ and effective mass parameter $a$ for random noisy initial conditions and fixed pump intensity. For fixed $\gamma_c$ the number of dark solitons increases with increasing effective mass (decreasing parameter $a$). When the polariton mobility is reduced at smaller $a$, more phase defects in the initial noise survive. At $R=1$, dark solitons are formed at each pump valley [Fig.~\ref{f2}(c)]. Fig.~\ref{f2}(d) shows a case for larger effective mass where only few dark solitons are formed. For $R\rightarrow0$ no dark solitons survive. For fixed $a$, $a=1$ for instance, the spatial correlations inside the condensate are more pronounced if the polaritons have a longer lifetime (smaller $\gamma_c$), resulting in the formation of less dark solitons. In summary, the less mobile the polaritons and the shorter their lifetime, the larger the number of dark solitons formed from initial noise. The number of dark solitons further depends on pump intensity \cite{Liew-prb-2015} as shown in Fig.~\ref{f2}(b) for fixed $\gamma_c$ and $a$.

\begin{figure} %[htbp]
\includegraphics[width=0.95\columnwidth]{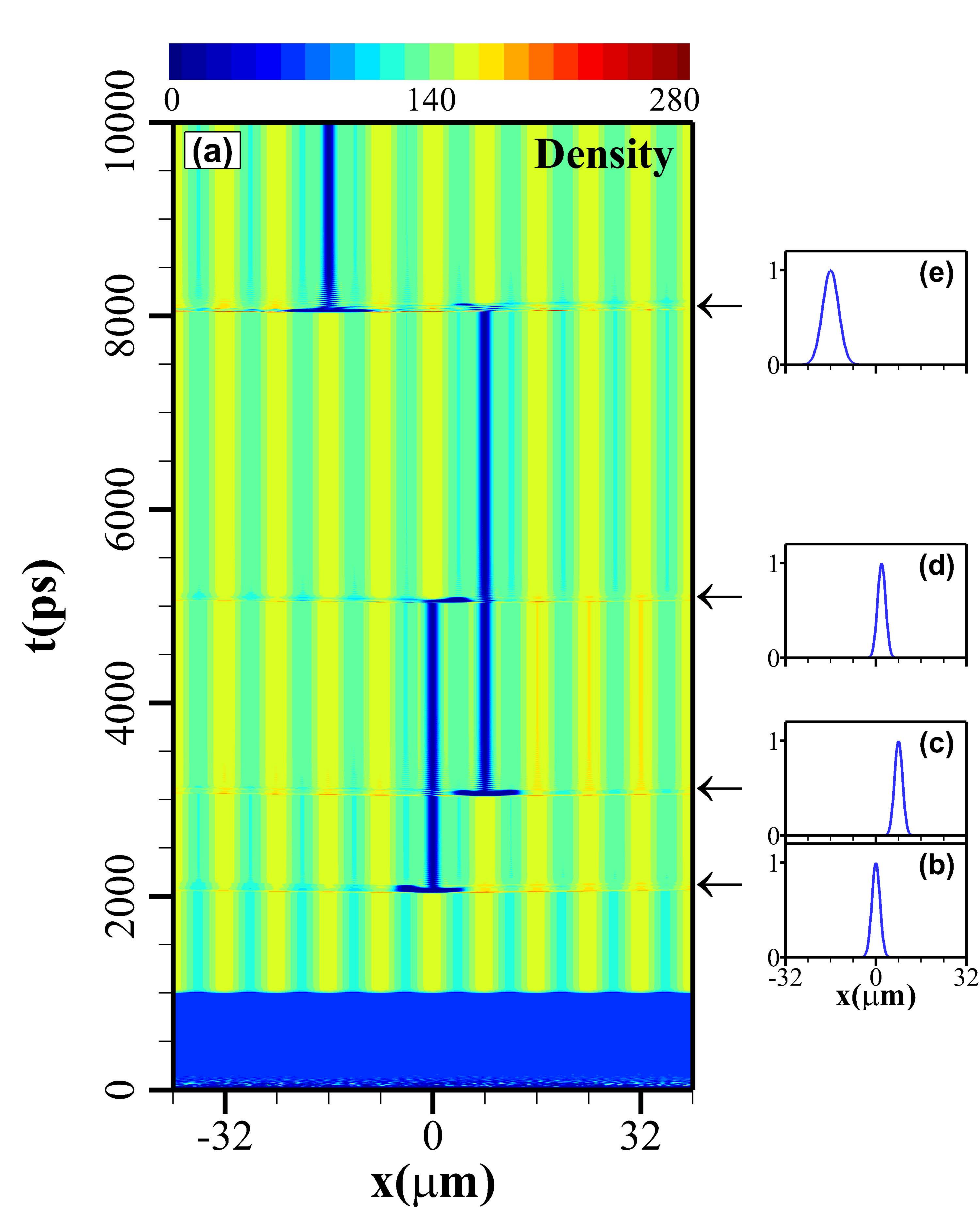}
\caption{(Color online) Controlled excitation and switching dynamics of dark solitons. (a) Time evolution of the condensate density under homogeneous ($t<1000\,\mathrm{ps}$) and periodic ($t\geq1000\,\mathrm{ps}$) excitation at $P=30$. (b)-(e) Normalized spatial intensity profiles of coherent pulse used for creation and annihilation of dark solitons launched at different times and spatial positions (b) $x=0\,\mathrm{\mu m}$, (c) $x=8\,\mathrm{\mu m}$, (d) $x=2\,\mathrm{\mu m}$, (e) $x=-16\,\mathrm{\mu m}$.}\label{f3}
\end{figure}

In experiments on microcavity polaritons, initial conditions are always noisy. Therefore it is difficult to excite a pure periodic solution, a single dark soliton, or a dark soliton at a specific pump valley. In Fig.~\ref{f3} we present an approach where we use the fact that dark solitons are unstable under homogeneous excitation \cite{Xue-2014-prl}. We use the excitation scenario illustrated in Fig.\ref{f1}(a) with a time delay (here one nanosecond) between the two pump beams. The beam arriving first excites a homogeneous condensate solution [$t<1000\,\mathrm{ps}$ in Fig.~\ref{f3}(a)], which rapidly forms from initial noise \cite{Matuszewski-prb-2014}. Then, the second beam arrives and interferes with the first beam, periodically modulating the pump profile. This periodic modulation is also transferred to the condensate for which a periodic solution forms without generation of any dark solitons [$1000\,\mathrm{ps}\leq t<2000\,\mathrm{ps}$ in Fig.~\ref{f3}(a)]. After initialization of this periodic state, an additional coherent pulse with $3\mathrm{\mu m}$ width and $80\,\mathrm{ps}$ duration [Fig.~\ref{f3}(b)] is used to create a dark soliton. Previously it was reported that such a coherent pulse with Gaussian shape suffers a depletion at its center and the condensate forms a ring shape because of propagation of the condensate away from the source \cite{Dominici-ncom-2015}. In Fig.~\ref{f3}(a) a region with low polariton density is generated near $x=0$ after the injection of the coherent pulse at $t=2000\,\mathrm{ps}$. At the same time the phase coherence is disturbed and a dark soliton is created. With another coherent pulse at $t=3000\,\mathrm{ps}$ we create another dark soliton in a neighboring pump valley. With the same approach further dark solitons can be created. As the dark solitons only survive in the pump valleys, a coherent pulse injected at the center of a pump peak between the two existing dark solitons (for example at $x=4\,\mathrm{\mu m}$) can be used to simultaneously annihilate both neighboring dark solitons at $x=0$ and $x=8\,\mathrm{\mu m}$ (not shown). If the coherent pulse is launched closer to one of the two solitons [at $x=2\,\mathrm{\mu m}$ in Fig.~\ref{f3}(d)], the nearest dark soliton at $x=0$ is annihilated whereas the other (at $x=8\,\mathrm{\mu m}$) survives [$5000\,\mathrm{ps}<t<8000\,\mathrm{ps}$ in Fig.~\ref{f3}(a)]. A broader coherent pulse with width of $6\,\mathrm{\mu m}$ [Fig.~\ref{f3}(e)] and $50\,\mathrm{ps}$ duration can be used to annihilate a dark soliton in a more distant pump valley and simultaneously create a new dark soliton where the broader pulse is injected.

\begin{figure} %[htbp]
\includegraphics[width=1.0\columnwidth]{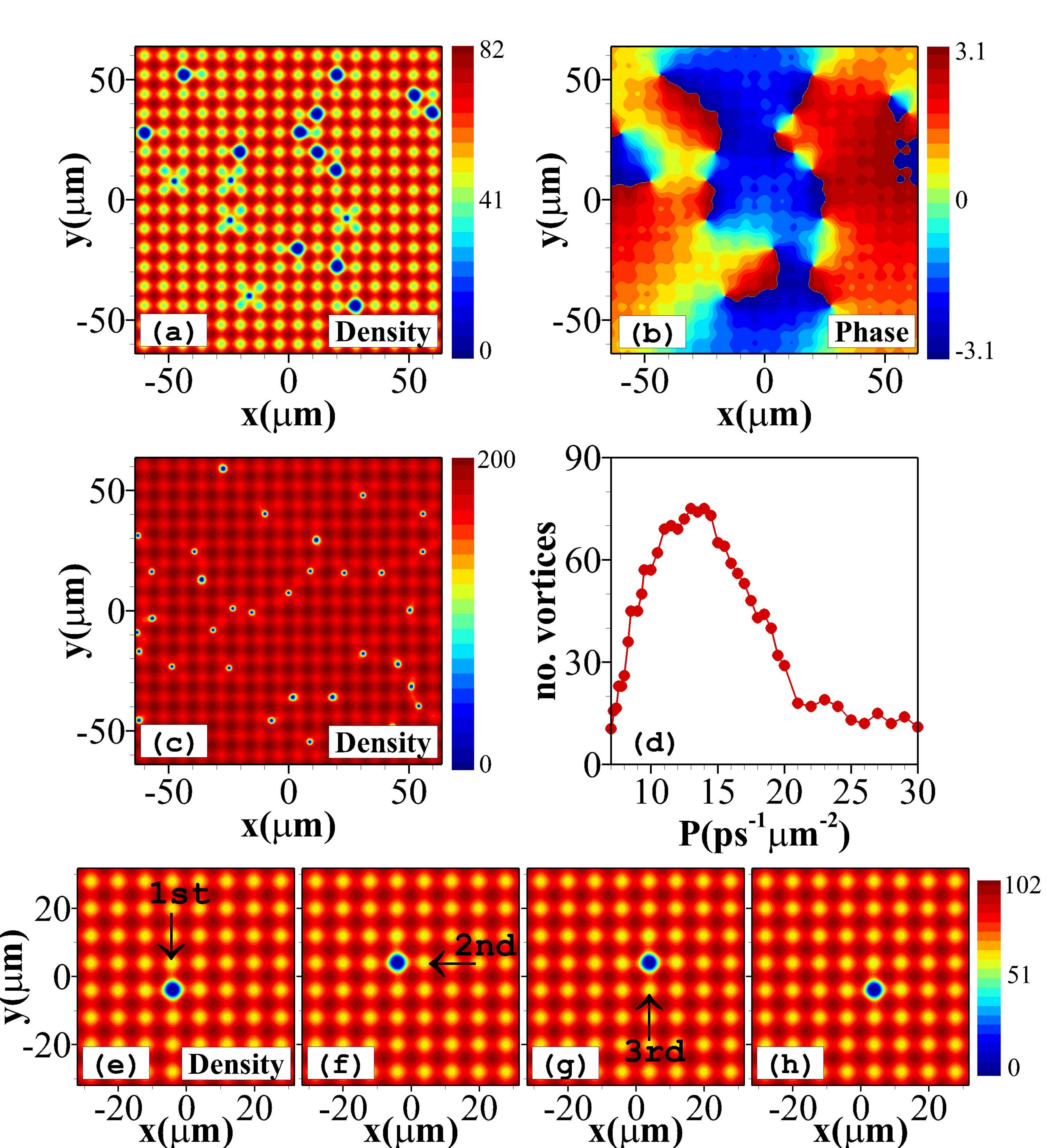}
\caption{(Color online) Controlling vortices with coherent pulses. Profiles of (a) density and (b) phase of condensate under 2D periodic pump excitation at $P=8$ showing vortices (phase defects) forming from the initial noise. (c) Density profile of condensate at $P=20$. (d) Dependence of number of vortices on pump intensity. (e)-(h) Manipulation of a vortex by a sequence of four coherent pulses at $P=10$. Snapshots of the density profiles are shown  at (e) $t=400\,\mathrm{ps}$, (f) $t=800\,\mathrm{ps}$, (g) $t=1200\,\mathrm{ps}$, and (h) $t=1400\,\mathrm{ps}$. A full movie is included in the Supplemental Material. Arrows indicate the target cell for vortex motion due to application of a coherent pulse.}\label{f4}
\end{figure}

\textit{2D excitation} -- Previously it was reported that in a two-dimensional system a homogeneous pump supports stable vortices in polariton condensates \cite{Liew-prb-2015}. In the present work we use excitation with a 2D periodic lattice given by $P_i(x,y)=P(\sin^2(\pi x/8)+\sin^2(\pi y/8))$ as illustrated in Fig.~\ref{f1}(a). For this type of excitation, we find two different types of vortices that form from initial noise as illustrated in Figs.~\ref{f4}(a) and (b). As shown in Fig.~\ref{f4}(d) the total number of vortices initially increases with increasing pump intensity [$P<14$], analogously to the scaling laws discussed in \cite{Liew-prb-2015}. For larger pump intensity, however, due to saturation effects the periodic condensate background is more similar to the homogeneous solution, Fig.~\ref{f4}(c).

Unlike the controlled creation of dark solitons by coherent pulses in the 1D case, it has proven more difficult to create a vortex in the 2D periodic condensate. We find, however, that coherent pulses can be used to re-locate an existing vortex from its original cell to a neighboring cell as shown in Fig.~\ref{f4} (a movie showing the dynamics is available in the Supplemental Material). In Figs.~\ref{f4}(e)-(h) we use a sequence of four pulses, each $3\,\mathrm{\mu m}$ in diameter and $80\,\mathrm{ps}$ in length. We find that a coherent pulse creates a reduced condensate density at its center. Applied next to an existing vortex this can be used to reduce the effective potential confining the vortex on one side. Such that the vortex is then released. Applying a series of pulses in different spatial positions, a vortex can be systematically moved around.

\begin{figure} %[htbp]
\includegraphics[width=1.0\columnwidth]{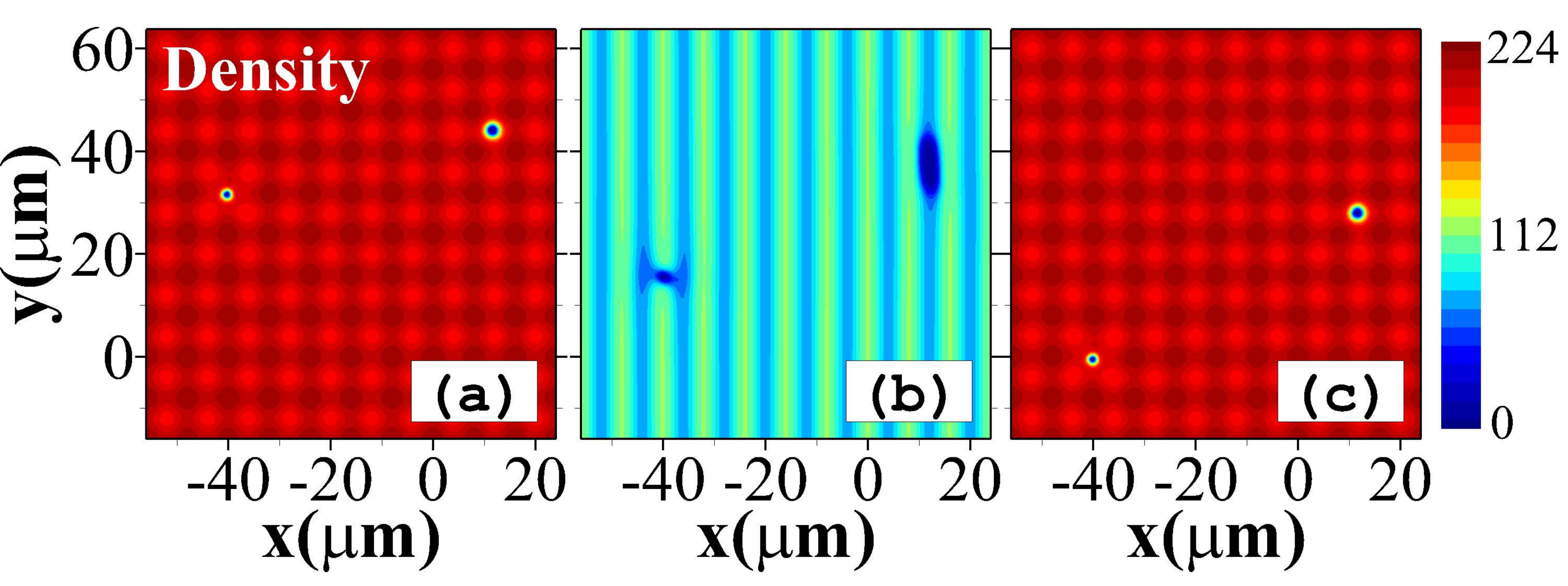}
\caption{(Color online) Controlling vortices through non-resonant excitation. Shown is the manipulation of vortices by switching the excitation from a two-dimensional lattice (with two vortices) in (a) to a stripe-like excitation in (b), setting the vortices into motion, back to a two-dimensional lattice confining the vortices in a different place in (c). The panels show snapshots of the condensate density in time for $P=22$ at (a) $t=400\,\mathrm{ps}$, (b) $t=800\,\mathrm{ps}$, and (c) $t=1200\,\mathrm{ps}$. A corresponding movie is included in the Supplemental Material.}\label{f5}
\end{figure}

Finally we would like to discuss that the motion of vortices in 2D can also be manipulated only modifying the source that generates the periodic reservoir density to a stripe-like source with $P_i(x,y)=P\sin^2(\pi x/8)$. In this configuration the vortices can move freely along the stripe direction. This can be used to control the position of vortices after creation on a 2D lattice. For example, consider a case where two vortices were created by a 2D periodic pump lattice [Fig.~\ref{f5}(a)]. If subsequently switching to a stripe-like pump profile [Fig.~\ref{f5}(b)], the two vortices move along the $y$ axis. When switching back to the 2D lattice at a later point in time [Fig.~\ref{f5}(c)] the two vortices are trapped in a different place. A movie is included in the Supplemental Material. Analogously, the vortices can be moved along the $x$-direction. However, we note that the direction of motion (forward or backward) and the velocity acquired by the vortices is not easy to predict in this simple scheme, as it is influenced by the phase distribution and the interactions between vortices.

To analyze the robustness of the phase defects studied here, we have investigated the influence of  fabrication-induced disorder on our results \cite{Savona-jp-2007}. A random disorder potential realistic for high-quality structures with a depth of $V_d\lesssim0.1meV$ \cite{Fraser-njp-2009} and a spatial correlation length of one micron was assumed. Our numerical results demonstrate that both 1D and 2D phase defects including their optical control are indeed quite robust and our results are not qualitatively altered by the presence of such realistic disorder.

\textit{Conclusion} -- To summarize, we found that when creating a polariton condensate with a spatially periodic pump source, phase defects are stabilized in both 1D and 2D. In 1D we demonstrate that dark solitons are either generated randomly from the initial noise triggering condensation or -- more importantly -- can be created on demand by a coherent light pulse. In 2D, we find two kinds of stable vortices which can also be optically manipulated and flexibly moved on a 2D lattice by use of coherent light pulses. We further demonstrate that vortices in 2D can also be manipulated using only non-resonant optical means. Our findings open up exciting possibilities to use dark solitons and vortices in polariton condensates for information storage and processing or quantum simulators.

This work was supported by the Deutsche Forschungsgemeinschaft (DFG) through the collaborative research center TRR 142. S.S. acknowledges support through the Heisenberg program of the DFG.

\end{document}